\documentclass[preprint,showpacs,preprintnumbers,amsmath,amssymb]{revtex4}

\usepackage{graphicx}
\usepackage{dcolumn}
\usepackage{bm}

\begin{document}

\preprint{}

\title{Two-Qubit Separability Probabilities and Beta Functions}

\author{Paul B. Slater}%
\email{slater@kitp.ucsb.edu}
\affiliation{%
ISBER, University of California, Santa Barbara, CA 93106\\
}%
\date{\today}

\begin{abstract}
Due to recent important 
work of {\.Z}yczkowski and Sommers ({\it J. Phys. A} 
{\bf{36}}, 10115 [2003] and {\bf{36}}, 10083 [2003]), 
exact formulas are available (both in terms of
the Hilbert-Schmidt and Bures metrics) for
the $(n^2-1)$-dimensional and $(\frac{n(n-1)}{2}-1)$-dimensional 
volumes of the complex and 
real $n \times n$ density matrices.
However, no comparable formulas are available for the volumes 
(and, hence, probabilities) of 
various separable
subsets of them. 
We seek to clarify this situation for the Hilbert-Schmidt metric for 
the simplest possible case of $n=4$, that is,
the two-qubit systems. Making use of the density matrix ($\rho$) 
parameterization
of Bloore ({\it J. Phys. A} {\bf{9}}, 2059 [1976]),
we are able to 
reduce each of the real and complex volume problems to the calculation
of a {\it one}-dimensional integral, the single relevant variable 
being a certain ratio of diagonal entries,
$\nu = \frac{\rho_{11} \rho_{44}}{\rho_{22} \rho_{33}}$. The associated integrand in each 
case is the
product of a {\it known} (highly oscillatory near $\nu=1$) 
jacobian and a certain {\it unknown} univariate function, which 
our extensive
numerical (quasi-Monte Carlo)  
computations indicate is very closely proportional to an
(incomplete) beta function $B_{\nu}(a,b)$, with $a=\frac{1}{2}, 
b=\sqrt{3}$ in the real case, and $a=\frac{2 \sqrt{6}}{5}, 
b =\frac{3}{\sqrt{2}}$ in the complex case. 
Assuming the full applicability of these specific incomplete beta
functions, we undertake separable volume calculations.
\end{abstract}


\maketitle
\section{Introduction}
In a pair of major, skillful papers, making use of the theory of
 random matrices \cite{random}, Sommers and \.Zyczkowski were able to
derive explicit formulas for the volumes occupied by the 
$d= (n^2-1)$-dimensional convex set of $n \times n$ (complex) 
density matrices (as well as the $d=\frac{(n-1)(n+2)}{2}$-dimensional 
convex set of real (symmetric) $n \times n$ density matrices),
both in terms of the Hilbert-Schmidt (HS) metric \cite{szHS} --- inducing the flat, Euclidean geometry --- and 
the Bures metric \cite{szBures} (cf. \cite{szMore}).
Of course, it would be of obvious 
considerable quantum-information-theoretic 
interest in the cases that $n$ is a composite
number, to also obtain HS and Bures volume 
formulas restricted to those states that
are separable --- the sum of 
product states --- in terms of some factorization of $n$ \cite{ZHSL}. 
Then, by taking ratios --- employing these Sommers-\.Zyczkowski 
results --- one would obtain corresponding 
separability {\it probabilities}.
(In an {\it asymptotic} regime, in which the dimension of the state space
grows to infinity, Aubrun and Szarek 
recently concluded \cite{aubrun} that for
qubits and larger-dimensional particles, the proportion of the
states that are separable is superexponentially small in the dimension
of the set.)

In particular, again for the 15-dimensional complex 
case, $n=4 = 2 \times 2$, numerical
evidence has been adduced 
that the Bures volume of separable states is
(quite elegantly) $2^{-15} (\frac{\sqrt{2}-1}{3}) \approx 
4.2136  \cdot 10^{-6}$ \cite[Table VI]{slaterJGP} 
and the HS volume
$(5 \sqrt{3})^{-7} 
\approx 2.73707 \cdot 10^{-7}$ \cite[eq. (41)]{slaterPRA}. 
Then, taking ratios (using the corresponding Sommers-\.Zyczkowski results), 
we have the derived conjectures that the Bures separability
 probability
is $\frac{1680 (\sqrt{2}-1)}{\pi^8} \approx 0.0733389$ 
and the HS one, 
considerably larger, $\frac{2^2 \cdot 3 \cdot 7^2 \cdot 11 \cdot 13 \sqrt{3}}{5^4 \pi^{6}} \approx 0.242379$ \cite[eq. (43), but misprinted as $5^3$ not 
$5^4$ there]{slaterPRA}.
(Szarek, Bengtsson and \.Zyczkowski --- motivated by the numerical 
findings of \cite{slaterPRA,slaterChinese} --- have recently 
formally demonstrated 
 ``that the probability to find a random state to be separable equals 2 times the probability to find a random boundary state to be separable, provided the random states are generated uniformly with respect to the Hilbert-Schmidt (Euclidean) distance. An analogous property holds for the set of positive-partial-transpose states for an arbitrary bipartite system'' \cite{sbz} 
(cf. \cite{innami}). These authors 
also noted \cite[p. L125]{sbz} that ``one could try to obtain similar
results for a general class of multipartite systems''. In this 
latter vein, preliminary
numerical analyses of ours have given some [but certainly not yet conclusive] 
indication that for the {\it three}-qubit
{\it tri}separable states, there may be an analogous probability ratio of
6 --- rather than 2.)

However, the analytical derivation of (conjecturally) exact
formulas for these HS and Bures (as well as other, such as the Kubo-Mori
\cite{petz1994} and Wigner-Yanase 
\cite{wigneryanase,slaterPRA}) 
separable volumes has seemed quite remote --- the only analytic 
progress to report so far being 
certain exact formulas
when the number of dimensions of the 15-dimensional space of $4 \times 4$ 
density matrices has been severely 
curtailed (nullifying or holding 
constant {\it most} of the 15 parameters) to $d \leq 3$ 
\cite{pbsJak,pbsCanosa} (cf. \cite{slaterC}).
Most strikingly, in this research direction,
in \cite[Fig. 11]{pbsCanosa}, we were able to find 
a highly interesting/intricate (one-dimensional) continuum 
($-\infty < \beta <\infty$) of two-dimensional 
(the associated 
parameters being $b_{1}$, the mean, and $\sigma_{q}^2$,   
the variance of the Bell-CHSH observable) 
HS separability
probabilities, in which the {\it golden ratio} \cite{livio} was 
featured --- serving to demarcate different separability regimes --- among 
other items. (The associated 
HS volume element --- $\frac{1}{32 \beta (1+\beta)} 
d \beta d b_{q} d \sigma^2_{q}$ --- is
independent of $b_{1}$ and $\sigma_{q}^2$ in this 
three-dimensional scenario.)
Further, in \cite{pbsJak}, building upon work of 
Jak\'obczyk and Siennicki \cite{jak}, we obtained a 
remarkably wide-ranging variety of exact HS 
separability ($n=4, 6$) and PPT (positive partial transpose) 
($n=8, 9, 10$) probabilities based on 
{\it two}-dimensional sections of sets of 
(generalized) Bloch vectors corresponding to $n \times n$ 
density matrices.

Nevertheless, 
computations for the {\it full} $d=9$ and/or $d =15$, $n=4$ real and complex 
two-qubit scenarios
are quite daunting --- due to the numerous
separability constraints at work, some being active [binding] 
in certain regions and 
in complementary regions, inactive [nonbinding]. 
``The geometry of the $15$-dimensional set of separable states of two
qubits is not easy to describe'' \cite[p. L125]{sbz}.
We seek to make substantial progress in these directions here, and, in fact, 
prove able to recast both these problems within {\it one}-dimensional 
frameworks.

We accomplish this dimensional reduction 
through the use of 
the (quite simple) form of parameterization of the density matrices
put forth by Bloore \cite{bloore,slaterJPAreject} some thirty years ago. 
(Of course, there are a number of
other possible 
parametrizations \cite{kk,byrd,sudarshan,vanik,fano,scutaru,stan}, 
several of 
which we have also utilized in various studies \cite{slaterA,slaterqip} 
to estimate volumes of 
separable states. Our greatest progress at this stage, 
in terms of increasing dimensionality,  though, has been achieved
with the Bloore parameterization --- due to a certain 
computationally attractive feature of it, allowing one to 
{\it decouple} diagonal and non-diagonal parameters --- which is described 
in sec.~\ref{sc1}.)
\subsection{Outline of paper}
In sec.~\ref{sc1} immediately below, 
we describe the  Bloore density matrix parameterization. Then,
we present in sec.~\ref{redsect} the specific one-dimensional 
integration formulas
we have obtained for the  real and complex HS 
separable qubit-qubit volumes using the Bloore parameterization.
The integrands in each of these cases 
are the product of a {\it known} jacobian function and a
heretofore {\it uncharacterized} function.
In sec.~\ref{estimation}, we detail the extensive numerical
(quasi-Monte Carlo) procedures employed to estimate these
unknown  functions.
Then, in sec.~\ref{findings},
we demonstrate --- quite unanticipatedly --- that 
our estimates of these functions 
over the unit interval are remarkably 
well-fitted (up to proportionality constants) by certain 
specific incomplete
beta functions. In the complex case, we can perform the indicated
separable-volume integration exactly, but only numerically in the real case.
In sec.~\ref{conclusions}, we give some concluding remarks.

In any case, it appears that further research 
is called for, 
to formally establish or appropriately qualify 
the role of the incomplete beta function 
in the determination of the real and complex two-qubit Hilbert-Schmidt 
{\it separable} volumes.

\section{Bloore parameterization of density matrices} \label{sc1}
The main presentation of Bloore \cite{bloore} 
was made in terms of the $3 \times 3$ ($n=3$)
density matrices. It is clearly easily extendible to cases
$n >  3$.
The fundamental idea is to scale the off-diagonal elements $(\rho_{ij}, 
i \neq j)$ 
of the density matrix in terms of the square roots of the diagonal 
entries ($\rho_{ii}$). That is, one sets (introducing the new [Bloore] 
variables $z_{ij}$),
\begin{equation}
\rho_{ij} = \sqrt{\rho_{ii} 
\rho_{jj}} z_{ij}.
\end{equation}
 This allows the determinant of $\rho$ (and analogously 
all its
principal minors) to be expressible as the {\it product}
($|\rho| = A_{1} A_{2}$) of
two factors, one ($A_{1} = \Pi_{i=1}^{4} \rho_{ii}$)  
of which is itself simply the product of 
(nonnegative) 
diagonal entries ($\rho_{ii}$). 
In the real $n=4$ case under investigation here --- we have
\begin{equation} \label{B}
A_{2}= \left(z_{34}^2-1\right) z_{12}^2+2 \left(z_{14}
   \left(z_{24}-z_{23} z_{34}\right)+z_{13}
   \left(z_{23}-z_{24} z_{34}\right)\right)
   z_{12}-z_{23}^2-z_{24}^2-z_{34}^2+
\end{equation}
\begin{displaymath}
z_{14}^2
   \left(z_{23}^2-1\right)+ z_{13}^2
   \left(z_{24}^2-1\right)+2 z_{23} z_{24} z_{34}+2 z_{13}
   z_{14} \left(z_{34}-z_{23} z_{24}\right)+1,
\end{displaymath}
involving (only) the $z_{ij}$'s ($i > j$), where $z_{ji}=z_{ij}$ 
\cite[eqs. (15), (17)]{bloore}.
Since, clearly, the factor $A_{1}$ is positive in all nondegenerate cases 
($\rho_{ii} > 0$),
one can --- by only analyzing $A_{2}$ --- essentially 
ignore the diagonal entries, and thus reduce by ($n-1$) the
dimensionality of the problem of finding nonnegativity 
conditions to impose on $\rho$.
This is the feature we have sought to maximally 
exploit above. A fully analogous 
decoupling property holds in the complex case.

It is, of course, necessary and sufficient for $\rho$ to serve
as a density matrix (that is, an Hermitian, nonnegative definite, trace
one matrix) that all its principal minors be nonnegative 
\cite{horn}.
The (necessary --- but not sufficient) 
condition --- quite natural in the Bloore 
parameterization --- that all the principal $2 \times 2$ minors be
nonnegative requires simply that $-1 \leq z_{ij} \leq 1, i \neq j$. The 
joint conditions that {\it all} the principal minors be nonnegative are not as
readily apparent. But for the 9-dimensional {\it real} 
case $n=4$ --- that is, $\Im(\rho_{ij})=0$ --- we have been able to obtain 
one such set,
using the Mathematica implementation of the {\it cylindrical
algorithm decomposition} (CAD) \cite{cylindrical}.
(The set of solutions of any system of real algebraic equations
and inequalities can be decomposed into a finite number of
``cylindrical'' parts \cite{strzebonski}.)
Applying it, we were able to express the 
conditions that 
an arbitrary  9-dimensional $4 \times 4$ real density matrix 
$\rho$ must fulfill.
These took the form,
\begin{equation} \label{limits}
 z_{12}, z_{13}, z_{14} \in [-1,1],
 z_{23} \in [Z^-_{23},Z^+_{23}],
 z_{24} \in [Z^-_{24},Z^+_{24}],
 z_{34} \in [Z^-_{34},Z^+_{34}],
\end{equation}
where
\begin{equation}
Z^{\pm}_{23} =z_{12} z_{13} \pm \sqrt{1-z_{12}^2} \sqrt{1-z_{13}^2} , 
Z^{\pm}_{24} =z_{12} z_{14} \pm \sqrt{1-z_{12}^2} \sqrt{1-z_{14}^2} ,
\end{equation}
\begin{displaymath}
Z^{\pm}_{34} = \frac{z_{13} z_{14} -z_{12} z_{14} z_{23} -z_{12} z_{13} z_{24} +z_{23}
z_{24} \pm s}{1-z_{12}^2},
\end{displaymath}
and
\begin{equation}
s = \sqrt{-1 +z_{12}^2 +z_{13}^2 -2 z_{12} z_{13} z_{23} +z_{23}^2}
\sqrt{-1 +z_{12}^2 +z_{14}^2 -2 z_{12} z_{14} z_{24} +z_{24}^2}.
\end{equation}
Making use of these results, we were able to confirm {\it via} exact 
symbolic integrations, 
the (formally demonstrated) 
result of \.Zyczkowski and Sommers
\cite{szHS} that the HS volume of the {\it real} 
two-qubit ($n=4$) states is
$\frac{\pi^4}{60480} \approx 0.0016106$.
(We could also verify this  through a somewhat [superficially, at least] 
different
Mathematica computation, using the implicit integration feature
first 
introduced in version 5.1. That is, the only integration limits employed were
that $z_{ij} \in [-1,1], i \neq j$ --- {\it broader} than those 
yielded by the CAD given by 
(\ref{limits}) --- while the Boolean constraints were now imposed that 
the determinant of $\rho$ and {\it one} [all that is needed to ensure
nonnegativity] of its principal $3 \times 3$ minors be nonnegative.)
\subsection{Determinant of the Partial Transpose}
However, when we tried to combine these 
CAD integration limits (\ref{limits}) 
with
the (Peres-Horodecki \cite{asher,michal,bruss} $n=4$) 
 separability constraint that the determinant ($A_{3} =|\rho_{PT}|$) 
of the partial
transpose of $\rho$ be nonnegative \cite[Thm. 5]{ver}, 
we exceeded the memory availabilities of our workstations.
In general, the term $A_{3}$ --- unlike the earlier term $A_{2}$ --- unavoidably 
involves 
the diagonal entries ($\rho_{ii}$), so the
dimension of the accompanying integration problems must increase, 
it would seem, we initially thought  --- 
in the $9$-dimensional real  $n=4$ case from six to nine.
\subsubsection{Role of {\it univariate} ratio of diagonal entries}
However, we then noted that, in fact, the dimensionality 
of the required integrations for the separable volumes must only
essentially be increased by one (rather than three) from that for 
the total volumes, since 
$A_{3}$ turns out to be (aside from the necessarily nonnegative 
factor of $A_{1}$, 
which we can ignore) 
expressible solely in terms of 
 the (six, in the real case) distinct
$z_{ij}$'s 
and the  ({\it univariate}) ratio
\begin{equation} \label{secondratio}
\nu = \frac{\rho_{11} \rho_{44}}{\rho_{22} \rho_{33}}.
\end{equation}
(Numerical probes of ours demonstrated that $\nu$ is {\it not}
a local invariant of two-qubit mixed states, in the sense of Makhlin
\cite{makhlin}.)
We, then, have 
\begin{equation}
A_{3} \equiv |\rho_{PT}| = 
 A_{1}  \Big(-z_{14}^2 \nu^2+2 z_{14} \left(z_{12} z_{13}+z_{24}
   z_{34}\right) \nu ^{3/2} + s \nu
+2 z_{23}
   \left(z_{12} z_{24}+z_{13} z_{34}\right) \nu^{1/2} -z_{23}^2 \Big),
\end{equation}
where 
\begin{displaymath}
s= \left(z_{34}^2-1\right)
   z_{12}^2-2 \left(z_{14} z_{23}+z_{13} z_{24}\right)
   z_{34} z_{12}-z_{13}^2+z_{14}^2
   z_{23}^2+\left(z_{13}^2-1\right) z_{24}^2-z_{34}^2-2
   z_{13} z_{14} z_{23} z_{24}+1.
\end{displaymath}

$A_{3}$ is, thus,  a {\it quartic/biquadratic} polynomial in terms of $\sqrt{\nu}$
(cf. \cite{wang,sudarshan}). 
(Clearly, the difficulty of the 
two-qubit separable-volume problem under study here
is strongly tied to the high [fourth] degree of $A_{3}$ in $\sqrt{\nu}$. 
By setting either $z_{14}=0$ or $z_{23}=0$, the degree of $A_{3}$
can be reduced to 2 (cf. \cite{slaterJPAreject}).)
In the {\it complex} case, $A_{3_{complex}}$ --- which we do not explicitly
present here --- also assumes the form of a
quartic polynomial in $\sqrt{\nu}$. So one must deal, in such a  setting, with
13-dimensional integration
problems rather than 15-dimensional ones.
\subsubsection{attempted seven-fold exact integration}
The problem of determining the separable volumes 
can, thus,  be seen to hinge on (in the
real case), a {\it seven}-fold
integration involving the six (independent) $z_{ij}$'s and $\nu$.
However, such requisite integrations, allowing $\nu$ to vary (or even holding
$\nu$ constant at various values, such as $\nu =1$, thus, reducing to six-fold integrations), 
did not appear --- rather frustratingly, we must admit  --- to be 
exactly/symbolically  
performable (using version 5.2 of 
Mathematica).

\section{Reduction to {\it One}-Dimensional Problems} \label{redsect}
Making use of the 
``Bloore parameterization'' \cite{bloore} (sec.~\ref{sc1} 
immediately above) 
of density matrices, which allows 
the decoupling of diagonal entries from non-diagonal entries in certain 
relevant determinant calculations, one can show that the problem of
computing the 15-dimensional volume ($V_{sep/complex}$) 
of the {\it separable} two-qubit systems is 
reducible to a {\it one}-dimensional integration of the form,
\begin{equation} \label{Vcomplex}
V_{sep/complex} = 2 \int_{0}^{1} Jac_{complex}(\nu) F_{complex}(\nu) d \nu.
\end{equation}
(We measure volume in terms of the Euclidean/Hilbert-Schmidt/Frobenius
norm, and slightly modify our notation in \cite{univariate}, to indicate
that we have changed from the variable $\mu$ used there to $\nu \equiv 
\mu^2$ here. 
The variable $\nu$ --- as noted earlier (\ref{secondratio}) --- is 
simply 
a specific 
ratio $\frac{\rho_{11} \rho_{44}}{\rho_{22} \rho_{33}}$ of diagonal entries $\rho_{ii}$ of the corresponding $4 \times 4$ density matrices
$\rho$.)

\begin{figure}
\includegraphics{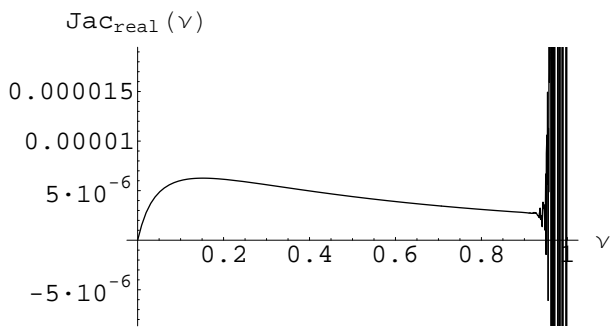}
\caption{\label{fig:jacreal}Plot of the jacobian function
$Jac_{real}(\nu)$, given by (\ref{Jacreal})}
\end{figure}

\begin{figure}
\includegraphics{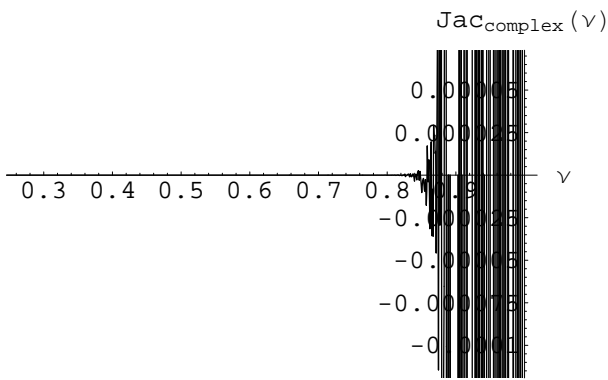}
\caption{\label{fig:jaccomplex}Plot of the jacobian function
$Jac_{complex}(\nu)$, given by (\ref{Jaccomplex})}
\end{figure}

Similarly, the 9-dimensional Hilbert-Schmidt volume of the 
separable real
density matrices (those with entries restricted to the real numbers) can 
be expressed as
\begin{equation} \label{Vreal}
V_{sep/real} = 2 \int_{0}^{1} Jac_{real}(\nu) F_{real}(\nu) d \nu.
\end{equation}
The two (highly oscillatory near $\nu=1$) 
jacobian functions ($Jac_{real}(\nu)$ and $Jac_{complex}(\nu)$) are
both explicitly known \cite[sec. III.B]{univariate}, 
that is (Fig.~\ref{fig:jacreal}),
\begin{equation} \label{Jacreal}
Jac_{real}(\nu) = \frac{\nu ^{3/2} \left(12 \left(\nu  (\nu +2) \left(\nu
   ^2+14 \nu +8\right)+1\right) \log \left(\sqrt{\nu
   }\right)-5 \left(5 \nu ^4+32 \nu ^3-32 \nu
   -5\right)\right)}{3780 (\nu -1)^9}
\end{equation}
and (Fig.~\ref{fig:jaccomplex})
\begin{equation} \label{Jaccomplex}
Jac_{complex}(\nu) = 
-\frac{\nu ^3}{3603600 (\nu -1)^{15}}(h_{1}+h_{2}),
\end{equation}
where
\begin{displaymath}
h_{1} = 363 \nu ^7+9947 \nu ^6+48363 \nu ^5+42875 \nu ^4-42875 \nu
   ^3-48363 \nu ^2-9947 \nu -363;
\end{displaymath}
\begin{displaymath}
h2=-140 \left(\nu ^7+49 \nu ^6+441 \nu ^5+1225 \nu ^4+1225
   \nu ^3+441 \nu ^2+49 \nu +1\right) \log \left(\sqrt{\nu
   }\right).
\end{displaymath}

We obtained the jacobian functions $Jac_{real}(\nu)$ and
$Jac_{complex}(\nu)$, given in (\ref{Jacreal}) and (\ref{Jaccomplex}),
by transformations of, say, $\rho_{33}$ to the $\nu$ variable
(and subsequent two-fold exact integrations over $\rho_{11}$ and
$\rho_{22}$) of
the original (three-dimensional) jacobians,
involving the diagonal entries, for the Bloore parameterizations. These
original
jacobians were of the form
$(\Pi_{i=1}^{4} \rho_{ii})^k$
with
$k=\frac{3}{2}$ in the real case, and $k=3$, in the complex case.
(Of course, by the unit trace condition,
we must have $\rho_{44}=1-\rho_{11}-\rho_{22}-\rho_{33}$.)

The (only) unknowns in our two separable-volume-computation problems 
((\ref{Vcomplex}) and (\ref{Vreal})) are, then,
the functions $F_{real}(\nu)$ and $F_{complex}(\nu)$.
In our preprint \cite{univariate}, we reported our initial 
numerical (quasi-Monte Carlo)
procedures to estimate these two centrally important functions 
(but in terms of the variable $\mu=\sqrt{\nu}$).
We have since continued these efforts, which we now detail in the following
section.
\section{Estimation of Unknown Univariate Functions} \label{estimation}
At an advanced stage of our numerical analyses, the initial results of 
which had been reported in \cite{univariate}, it appeared that 
it might be more efficacious to employ
$\nu = \frac{\rho_{11} \rho_{44}}{\rho_{22} \rho_{33}}$ as the
principal variable rather than 
$\mu = \sqrt{\frac{\rho_{11} \rho_{44}}{\rho_{22} \rho_{33}}}$.
(Thus, as previously noted, we have $\nu \equiv \mu^{2}$.)
So, our estimation ({\it uniform} sampling) 
procedures were originally 
designed in terms of $\mu$, rather than
$\nu$.

We had, in \cite{univariate}, begun 
proceeding along two fully parallel courses, one for the 9-dimensional
real two-qubit case and the other for the 15-dimensional complex case.
We sought those functions $f_{real}(\mu)$ and $f_{complex}(\mu)$, that we now see satisfy
the equivalences,
\begin{equation}
f_{real}(\sqrt{\nu}) \equiv F_{real}(\nu), \hspace{.2in}
f_{complex}(\sqrt{\nu}) \equiv F_{complex}(\nu),
\end{equation}
that would result from imposing the conditions that 
the expressions $A_{1}$, $A_{2}$ and $A_{3}$
(as well as a principal $3 \times 3$ minor of $\rho$), along with
their complex counterpart expressions, 
be {\it simultaneously} 
nonnegative. (The satisfaction of all these 
joint conditions ensures that we are dealing precisely
with {\it separable} $4 \times 4$ density matrices.) 
It was evident that the relation $f(\mu)=f(\frac{1}{\mu})$ must hold,
so we only numerically 
studied the range $\mu \in [0,1]$. Dividing this unit interval
into 2,000 equal nonoverlapping subintervals of length $\frac{1}{2000}$ 
each, 
we sought to estimate the $f(\mu)$'s at the 2,001 end points of these
subintervals.

This required ($\mu$ being alternately fixed at every single one 
of these  
end points for each set of $z_{ij}$'s) numerical
integrations in 6 and 12 dimensions. For this purpose, we utilized the
Tezuka-Faure (TF) quasi-Monte Carlo procedure \cite{giray1,tezuka}, 
we have extensively used in
our earlier studies of separability probabilities \cite{slaterJGP, slaterPRA} 
(cf. \cite{lovasz} for an apparently more efficient approach 
to estimating the Euclidean volume of {\it convex} bodies). 
For each of the 2,001 discrete,
equally-spaced values of $\mu$  we employed 
the same set of 611,500,000 Tezuka-Faure 
six-dimensional points in the real case and, similarly,  the 
 same
set of 549,500,000 twelve-dimensional points in the complex case.
(The Tezuka-Faure points are defined over unit hypercubes 
$[0,1]^{n}$, so in our computations, we transform the Bloore variables
accordingly and take into account the corresponding jacobians.)
\subsection{Close comparison with {\.Z}yczkowski-Sommers known values}
In the real case, our sample estimate of the
{\it known} Hilbert-Schmidt volume of (separable {\it plus} 
nonseparable) states \cite{szHS}, 
$\frac{\pi^4}{60480} \approx 0.0016106$ 
was smaller by only a factor of 0.999996. So, we would expect our 
companion estimates
of $f_{real}(\mu)$, at each of the 2,001 sampled points, 
to be roughly equally precise. 
(Let us note that $f_{real}(0) = f_{complex}(0)=0$. Statistical 
testing --- the use of confidence limits --- is not appropriate in
the Tezuka-Faure framework (cf. \cite{hong}).) 
In the complex case,
our estimate of the known
15-dimensional volume, $\frac{\pi^6}{851350500} \approx 
1.12925 \cdot 10^{-6}$ was larger only by a factor of 1.00009. 

As instances of specific values  (avoiding the 
necessity for 2,000 repetitions for each
point), based on independent analyses using
still larger numbers of TF-points,
we obtained estimates of $f_{real}(1) = F_{real}(1) = 
\frac{1610102144}{14046875} \approx 
114.62351, 
f_{real}(\frac{1}{2}) =  \frac{1040958844}{14046875} \approx 
74.10608$,  both based on 3,596,000,000 TF-points, and 
$f_{complex}(1) = F_{complex}(1) = 387.5080921$ and 
$f_{complex}(\frac{1}{2}) = 180.7173447$, both based on 
2,036,000,000 TF-points.
We have the {\it predicted} values  $G_{complex}(1) \approx 
387.486102$ and $G_{real}(1) \approx 114.6270015$. 
(Searches using the ``Plouffe's Inverter'' website 
http://pi.lacim.uqam.ca/eng/ did not readily yield any underlying
explanation of these values or a number of transformations of them.)

\section{Fitted incomplete beta functions} \label{findings}
\begin{figure}
\includegraphics{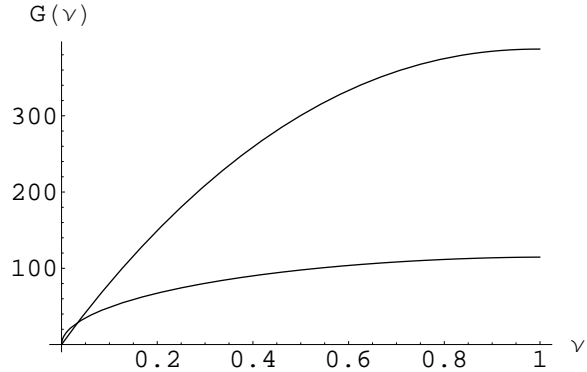}
\caption{\label{fig:Gfunctions}The two fitted scaled incomplete 
beta functions $G_{complex}(\nu)$ and the 
(lesser-valued at $\nu=1$) $G_{real}(\nu)$}
\end{figure}

\begin{figure}
\includegraphics{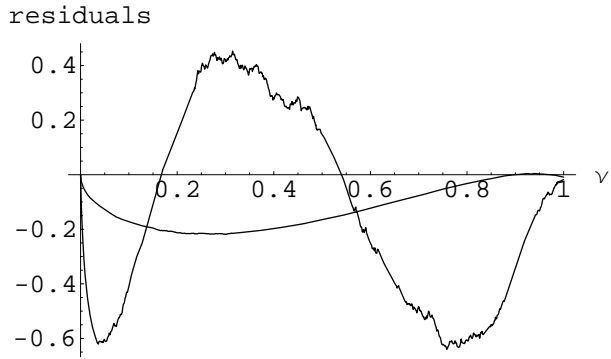}
\caption{\label{fig:residuals}Our numerical (interpolated) estimates 
(sec.~\ref{estimation}) of $F_{complex}(\nu)$ 
and $F_{real}(\nu)$ minus the values predicted by $G_{complex}(\nu)$ 
and $G_{real}(\nu)$. The more strongly fluctuating curve corresponds to the
complex case. Note the greatly reduced $y$-axis scale
{\it vis-{\` a}-vis} that of Fig.~\ref{fig:Gfunctions}, this observation
constituting
the basis for our central 
assertion that the $F(\nu)$'s are well fitted by the 
$G(\nu)$'s.}
\end{figure}

Numerical computations (detailed in sec.~\ref{estimation} above) --- provided 
us with  estimates of $F_{complex}(\nu)$ and $F_{real}(\nu)$ 
(though the sampling [quasi-Monte Carlo] procedure employed 
had been devised
in terms of the variable $\mu \equiv \sqrt{\nu}$ and counterpart 
functions $f_{complex}(\mu)$ and $f_{real}(\mu)$). 
We have been 
able to fit these results quite
well (Fig.~\ref{fig:residuals}) 
using (concave) functions  of the form (Fig.~\ref{fig:Gfunctions}),
\begin{equation} \label{candidateReal}
G_{real}(\nu) = \left(4+\frac{1}{5 \sqrt{2}}\right)
   B\left(\frac{1}{2},\sqrt{3}\right)^8 
B_{\nu }\left(\frac{1}{2},\sqrt{3}\right)
\end{equation}
and 
\begin{equation} \label{candidateComplex}
G_{complex}(\nu) = \left( \frac{100000000}{2
   \sqrt[3]{2}+\frac{10^{3/4}}{3^{2/3}}} \right)
B\left(\frac{2
   \sqrt{6}}{5},\frac{3}{\sqrt{2}}\right)^{14}
B_{\nu }\left(\frac{2
   \sqrt{6}}{5},\frac{3}{\sqrt{2}}\right).
\end{equation}
(Let us note that $\sqrt{3} \approx 1.73205, \frac{2 \sqrt{6}}{5} \approx 
0.979796$ and
$\frac{3}{\sqrt{2}} \approx 2.12132$.)
Here $B$ denotes the (complete) 
beta function, and $B_{\nu}$ the {\it incomplete} beta 
function \cite{handbook},
\begin{equation}
B_{\nu}(a,b) =\int_{0}^{\nu} w^{a-1} (1-w)^{b-1} d w.
\end{equation}
(The beta function itself, that is  $B(a,b) \equiv B_{1}(a,b)$,
played an important instrumental role in the original
formulation of string theory \cite[p. 6]{string}. 
For a more specific-still {\it incomplete} beta function role in string
theory, pertaining to the symmetric group $S_{3}$ and the 
modular group $M(2)$, see the review 
MR512916 of
A. O. Barut in the MathSciNet database [http://ams.rice.edu/mathscinet/] 
of a [somewhat obscure]
paper of M. Z{\u a}g{\u a}nescu \cite{zag1}.)

To obtain the two 
residual curves shown in Fig.~\ref{fig:residuals} --- upon which
we draw our central
conclusion that $F_{real}(\nu)$ and $F_{complex}(\nu)$
are well fitted by $G_{real}(\nu)$ and $G_{complex}(\nu)$,
respectively --- we interpolated the Tezuka-Faure points, using third-order
polynomials --- and then reparameterized the resulting curve in terms of $\nu$.
\subsection{separable volume {\it and} hyperarea estimations} \label{sve}
From the (exact) formulas of {\.Z}yczkowski and Sommers \cite{szHS}
for the Hilbert-Schmidt volumes of the real and complex $n \times n$ density
matrices to the case $n=4$, we know that the {\it total} volume of 
separable {\it and} 
nonseparable two-qubit systems is 
$\frac{\pi^6}{851350500} \approx 1.12925 \cdot 10^{-6}$ 
in the 15-dimensional (complex) case 
and $\frac{\pi^4}{60480} \approx 0.0016106$ in the 9-dimensional (real) case.

Also, from the results of {\.Z}yczkowski and Sommers \cite[eq. (6.5)]{szHS}, 
one can readily deduce that the 
ratio of boundary (14-dimensional) 
hyperarea to volume of the 15-dimensional convex set
of $4 \times 4$ density matrices is equal to $30 \sqrt{3}$, and
further \cite[eq. (7.9)]{szHS} 
that the corresponding (lesser) ratio for the 9-dimensional convex set 
of {\it real} $4 \times 4$ density matrices is $18 \sqrt{3}$. 
By the subsequent results of Szarek, Bengtsson and {\.Z}yczkowski 
\cite{sbz} --- which were 
motivated by certain numerical analyses of Slater
\cite{slaterPRA} --- we 
know the analogous hyperarea-volume ratios for the 15- and 9-dimensional 
{\it separable} subsets must be simply {\it twice} as large (that is, 
$60 \sqrt{3}$ and $36 \sqrt{3}$).

Using the proposed incomplete beta function fits 
 (\ref{candidateReal}) and (\ref{candidateComplex}),
we have attempted the evaluations of the two corresponding separable
volumes ((\ref{Vcomplex}) and (\ref{Vreal})), 
as well as separable (lower-dimensional) hyperareas, 
obtaining exact results in the complex case, 
but only numerical ones for the real scenario.
We succeeded in the complex 
case, using integration-by-parts, first integrating
$Jac_{complex}(\nu)$. (In the real case, an analogous initial integration
of $Jac_{real}(\nu)$ led to a much more complicated result, now involving
various hypergeometric functions. So, the integration by parts was stymied
there.) The exact result itself in the complex case 
(for which we thank M. Trott) 
was very lengthy 
(much too so to 
present here), but we could evaluate it to any given precision.
\subsubsection{complex case}
Using this {\it exact} formula,
we were able to obtain $V_{sep/complex} \approx  2.73827578 
\cdot 10^{-7}$, 
and again applying the {\.Z}yczkowski-Sommers \cite{szHS} and
Szarek-Bengtsson-{\.Z}ycskowski results \cite{sbz},
$P_{sep/complex} \approx 0.24248582$  and $H_{sep/complex} \approx 
0.0000142285$, all assuming the full applicability/validity of
(\ref{candidateComplex}).

We had previously hypothesized that $V_{sep/complex}=
(5 \sqrt{3})^{-7} \approx 2.73707 \cdot 10^{-7}$
\cite[eq. (41)]{slaterPRA} and
$P_{sep/complex}=  \frac{2^2 \cdot 3 \cdot 7^2 \cdot 11
\cdot 13 \sqrt{3}}{5^4 \pi^{6}}
\approx 0.242379$ \cite[eq. (43), but misprinted as $5^3$ not
$5^4$ there]{slaterPRA}. 
The analysis in \cite{slaterPRA} was based on 
400,000,000 quasi-Monte Carlo [Tezuka-Faure] 
points.
(Those points were 15-dimensional in nature
{\it vs.} the 12-dimensional ones used here.)
Additionally, each point there 
was employed only {\it once} for the Peres-Horodecki
separability test, while each point here is used in 2,000 such tests
(with $\mu$ ranging over [0,1]).
We had initially suspected that if we started 
checking the Peres-Horodecki criterion for successively larger values 
of  $\mu$, holding the set of $z_{ij}$'s given by a Tezuka-Faure point 
{\it fixed},  then if we reached one value of $\mu$ 
for which separability held, then
{\it all} higher values of $\mu$ (less than or equal to 1) 
would also yield separability. But this 
interestingly turned out
not to be invariably the case. So, it appeared that 
we needed to check the criterion 2,000 times
for every single 6-dimensional (real) or 12-dimensional 
(complex) TF-point.
\subsubsection{real case}
Since, as noted, exact integration-by-parts did not seem feasible 
in the real case,  we chose to expand
$G_{real}(\nu)$  in a
75-term power series about $\nu=0$, 
and performed {\it exact} integrations term-by-term.
The overall result can be expressed as 
$V_{sep/real} \approx 0.0007310253$.
Using  the various results of {\.Z}yczkowski and Sommers \cite{szHS}, 
and Szarek, Bengtsson and {\.Z}yczkowski \cite{sbz} 
detailed above,
we then immediately have the estimates $P_{sep/real} \approx 0.4538838$ 
for the separability {\it probability}
of the real $4 \times 4$ density matrices (markedly greater 
than in the complex 
case), and $H_{sep/real} \approx 
0.02279111$ for the hyperarea of the bounding 
8-dimensional hypersurface.
\section{Concluding Remarks} \label{conclusions}
In this study, using the Bloore parameterization of density matrices 
(\cite{bloore}, sec.~\ref{sc1}), 
we have shown that incomplete beta functions (sec.~\ref{findings}), or 
clearly quite close relatives
to them, appear to play important roles in the calculation of 
the Hilbert-Schmidt separable volumes of 
the 9-dimensional real and 15-dimensional complex qubit-qubit pairs.
However, there are still apparently systematic 
(sine-like) --- although quite small --- variations 
(Fig.~\ref{fig:residuals}) 
of the estimated function from the hypothesized one $G_{complex}(\nu)$ 
in the complex case, 
so we suspect that we may have 
possibly not yet fully explained this scenario.
So, to summarize, 
although we have developed here a rather compelling case for the
relevance of the incomplete beta functions, our evidence for this is
so far essentially empirical/numerical rather than theoretical.

The extension to qubit-{\it qutrit} pairs (and 
possibly 
higher-dimensional composite systems, $n>4$) 
of the {\it univariate}-function-strategy we have pursued above 
for the case of qubit-qubit pairs ($n=4$), seems problematical.
In the $n=4$ case, the analysis is
facilitated by the fact that it is sufficient that the {\it determinant} of
the partial transpose of a density matrix 
be nonnegative for the Peres-Horodecki separability
criterion to hold \cite[Thm. 5]{ver} \cite{augusiak}. 
{\it More} requirements than this single one
are needed in the qubit-qutrit
scenario --- even though the criterion of nonnegativity of the 
partial transpose  is still {\it both} necessary and 
sufficient for $6 \times 6$ density matrices. 
(In addition to the determinant, the 
leading minors and/or
the individual eigenvalues of the partial transpose of the 
$6 \times 6$ density matrix 
would need to be tested for nonnegativity, as well. 
Also the qubit-qutrit analogue
of the ratio ($\nu$) of diagonal entries, given by (\ref{secondratio}), 
 would have to be defined, if
even possible.)

In our earlier study \cite{slaterJPAreject}, we had also employed
the Bloore parameterization of the 
two-qubit (and qubit-qutrit) systems
to study the Hilbert-Schmidt (HS) separability probabilities of 
specialized systems of
{\it less} than full dimensionality.
We also reported there 
an effort to determine a certain {\it three}-dimensional
function (somewhat in contrast to the {\it one}-dimensional functions 
$F_{real}(\nu)$ and $F_{complex}(\nu)$ above, but for a rather similar
purpose) 
over the simplex of eigenvalues that would facilitate the
calculation of the 15-dimensional volume of the 
complex two-qubit systems in terms
of (monotone) metrics --- such as the Bures, Kubo-Mori, 
Wigner-Yanase,\ldots --- other than the (non-monotone 
\cite{ozawa}) Hilbert-Schmidt one considered here.
(The Bloore parameterization \cite{bloore} did not seem
immediately useful in this monotone metric context, since the {\it 
eigenvalues} of
$\rho$ are not explicitly expressed (cf. \cite{Dittmann}). 
Therefore, we had recourse in \cite{slaterJPAreject} to
the Euler-angle parameterization of density matrices of 
Tilma, Byrd and Sudarshan
\cite{sudarshan}, in which the eigenvalues of $\rho$ 
do, in fact, explicitly enter.)

So, it would 
seem to appear, initially at least, 
that the particular utility of the Bloore parameterization
in reducing the dimensionality of the problem of computing the Hilbert-Schmidt
separable 
volume of qubit-qubit pairs, of which we have taken advantage in this 
study, neither extends to higher-dimensional
Hilbert-Schmidt volumes ($n>4$) 
nor to monotone metric volumes of even qubit-qubit
pairs ($n=4$) themselves.

\begin{acknowledgments}
I would like to express gratitude to the Kavli Institute for Theoretical
Physics (KITP)
for computational support in this research 
and to Michael Trott for his advice on computing the separable
volumes (sec.~\ref{sve}).

\end{acknowledgments}

\bibliography{BetaResubmission}

\end{document}